\documentclass[twocolumn]{aastex701}

\newcommand{\Ha}{H$\alpha$}
\newcommand{\Hb}{H$\beta$}

\newcommand{\OII}{[O\,\textsc{ii}]}
\newcommand{\OIII}{[O\,\textsc{iii}]}

\newcommand{\SII}{[S\,\textsc{ii}]}
\newcommand{\NII}{[N\,\textsc{ii}]}
\newcommand{\sersic}{S\'{e}rsic}

\begin{document}

\title{Cold Gas Infall onto A Brightest Group Galaxy via A Gas-Rich Minor Merger}

\author[orcid=0000-0001-5105-2837,gname='Ming-Yang',sname=Zhuang]{Ming-Yang Zhuang}
\affiliation{Department of Astronomy, University of Illinois Urbana-Champaign, Urbana, 61801, IL, USA}
\email[show]{mingyang@illinois.edu}
\correspondingauthor{Ming-Yang Zhuang}

\author[orcid=0000-0002-4569-9009,gname=Savannah,sname=Shangguan]{Jinyi Shangguan}
\affiliation{Kavli Institute for Astronomy and Astrophysics, Peking University, Beijing, 100871, Beijing, People's Republic of China}
\affiliation{Department of Astronomy, School of Physics, Peking University, Beijing, 100871, Beijing, People's Republic of China}
\email{shangguan@pku.edu.cn}

\author[orcid=0000-0002-0210-946X,sname=Bian,gname=Yuan]{Yuan Bian}
\affiliation{Department of Astronomy, Xiamen University, Xiamen, 361005, Fujian, People's Republic of China}
\email{bianyuan@stu.xmu.edu.cn}

\author[orcid=0000-0003-1659-7035,gname=Yue, sname=Shen]{Yue Shen}
\affiliation{Department of Astronomy, University of Illinois Urbana-Champaign, Urbana, 61801, IL, USA}
\affiliation{National Center for Supercomputing Applications, University of Illinois Urbana-Champaign, Urbana, 61801, IL, USA}
\email{shenyue@illinois.edu}

\author[orcid=0000-0001-6947-5846,gname='Luis C.', sname=Ho]{Luis C. Ho}
\affiliation{Kavli Institute for Astronomy and Astrophysics, Peking University, Beijing, 100871, Beijing, People's Republic of China}
\affiliation{Department of Astronomy, School of Physics, Peking University, Beijing, 100871, Beijing, People's Republic of China}
\email{lho.pku@gmail.com}

\author[orcid=0000-0001-9953-0359,gname=Min, sname=Du]{Min Du}
\affiliation{Department of Astronomy, Xiamen University, Xiamen, 361005, Fujian, People's Republic of China}
\email{dumin@xmu.edu.cn}

\author[orcid=0000-0002-1605-915X,gname=Junyao, sname=Li]{Junyao Li}
\affiliation{Department of Astronomy, University of Illinois Urbana-Champaign, Urbana, 61801, IL, USA}
\email{junyaoli@illinois.edu}

\author[orcid=0000-0001-5017-7021,gname='Zhao-Yu', sname=Li]{Zhao-Yu Li}
\affiliation{Department of Astronomy, School of Physics and Astronomy, Shanghai Jiao Tong University, Shanghai, 200240, Shanghai, People's Republic of China}
\affiliation{Key Laboratory for Particle Astrophysics and Cosmology (MOE)/Shanghai Key Laboratory for Particle Physics and Cosmology, Shanghai, 200240, Shanghai, People's Republic of China}
\email{lizy.astro@sjtu.edu.cn}

\author[orcid=0000-0002-6593-8820,gname=Jing, sname=Wang]{Jing Wang}
\affiliation{Kavli Institute for Astronomy and Astrophysics, Peking University, Beijing, 100871, Beijing, People's Republic of China}
\email{jwang_astro@pku.edu.cn}

\submitjournal{ApJL}

\begin{abstract}

Dust and cold gas are not uncommon in nearby early-type galaxies (ETGs), and represent an important aspect of their evolution. However, their origin has been debated for decades. Potential sources include internal processes (e.g., mass loss from evolved stars), external mechanisms (e.g., minor mergers or cooling flows), or a combination of both. Gas-rich minor mergers have long been proposed as an important channel for cold gas fueling in both observations and simulations, but direct evidence of cold gas transportation via gas-rich minor mergers remains elusive, particularly in galaxy groups and clusters where environmental effects are prevalent. In this letter, we present the first unambiguous case of direct cold gas transportation onto a brightest group galaxy (BGG) at $z=0.25$, driven by an ongoing close-separation gas-rich minor merger with a mass ratio of $\sim1:56$. High-resolution JWST imaging reveals a heavily obscured, low-mass satellite that is barely visible at restframe optical wavelengths. Tidal stripping from this satellite deposits gas and dust onto the BGG, forming prominent $\sim$10 kpc dust lanes \textit{in situ}. Cosmological simulations indicate that such interactions preferentially occur in gas-rich satellites undergoing their first infall in highly eccentric orbits. Our results highlight the pivotal role of gas-rich minor mergers in replenishing cold gas reservoirs and shaping the evolution of central ETGs in galaxy groups.

\end{abstract}



\section{Introduction} 
Massive early-type galaxies (ETGs) have long been considered ``red and dead'' systems with little ongoing star formation and minimal cold gas reservoir \citep[e.g.,][]{Faber&Gallagher1976, Blanton&Moustakas2009}. However, extensive observational evidence over the past decades has revealed that dust and cold gas are not uncommon in nearby ETGs, representing an important aspect of their late-time evolution \citep[e.g.,][]{Welch&Sage2003ApJ, Morganti+2006, 2010MNRAS.409..500O}. The origin of these cold gas components has been debated for decades, with potential sources including internal processes (e.g., mass loss from evolved stars), external mechanisms (e.g., minor mergers or cooling flows), or a combination of both \citep[e.g.,][]{2001MNRAS.328..762E, 2003ARA&A..41..191M}. 

Cooling flows from the hot intracluster medium are considered the primary source of cold gas in galaxy clusters \cite[e.g.,][]{2003A&A...412..657S, 2019A&A...631A..22O}. In galaxy groups, however, the situation is less clear. Some galaxy groups exhibit similar properties to galaxy clusters, such as a hot intragroup medium with a cool core (CC) and ionized filaments, suggesting that rapid cooling flows may also contribute to their cold gas reservoir \citep{2017MNRAS.472.1482O, 2022A&A...666A..94O}. However, previous works have shown that while the fraction of CC brightest group galaxy (BGG) is relatively high ($\sim35\%$) in a high-richness subsample of the Complete Local-Volume Groups Sample, the cold gas fraction between CC and non-CC BGGs is not significantly different \citep{2015A&A...573A.111O, 2017MNRAS.472.1482O, 2018A&A...618A.126O}. Furthermore, no clear connection has been found between the presence of an X-ray detected intragroup and the molecular gas content in these objects \citep{2018A&A...618A.126O}. Similarly, the molecular gas detection rate does not show clear dependence on either large-scale or small-scale environment \citep{2011MNRAS.414..940Y, 2019MNRAS.486.1404D}. These results imply that cooling flows are unlikely to be the primary mechanism fueling cold gas reservoir in BGGs.

Gas-rich minor mergers have long been proposed to play an important role in fueling ETGs \cite[e.g.,][]{2009MNRAS.394.1713K, 2011MNRAS.411.2148K, 2019MNRAS.486.1404D}. Numerical simulations suggest that minor mergers are ubiquitous and contribute significantly to the cosmic star formation budget \cite[e.g.,][]{2014MNRAS.437L..41K, 2024MNRAS.527.6506B}. In particular, the reduced satellite velocity dispersion in galaxy groups compared to galaxy clusters lead to more frequent mergers between BGGs and their satellites \citep{2019MNRAS.485.2287B}. Additionally, ram pressure stripping is less efficient in galaxy groups with characteristic timescales of $\sim3$ Gyr due to the lower intragroup medium pressure and lower infalling velocity, resulting in a slower depletion of satellites' cold gas reservoir \citep{2021MNRAS.501.5073O, 2022A&ARv..30....3B}.

However, direct evidence for this process has remained elusive, with previous studies relying on indirect signatures such as gas-star misalignment, low gas-phase metallicity, extended and disturbed gas morphology, and frequent tidal features \cite[e.g.,][]{2006MNRAS.366.1151S, 2010MNRAS.409..500O, 2011MNRAS.414.2923K, 2013MNRAS.429..534D, 2020ApJ...905..154Y}. While recent interferometric observations have revealed disturbed atomic gas morphologies in HI-rich satellites at the outskirt of ETGs in the local Universe, the extent to which their cold gas can survive heating and mixing with the hot circumgalactic, intragroup, or intracluster medium and ultimately reach the center of ETGs remains highly uncertain \citep{2012MNRAS.422.1835S, 2016MNRAS.457..272D}. 

In this paper, we present the first direct evidence of cold gas transport through a gas-rich minor merger in the BGG of a massive galaxy group (dubbed CW-BGG-1) at $z=0.2475$ in the COSMOS field, with cold gas  transferred from an optically invisible, close satellite companion galaxy, CW-BGG-1-C. We describe the data and methods in Section~\ref{sec2}, present the structural, stellar, and gas properties of the system in Section~\ref{sec3}, compare our results with cosmological simulations and discuss the implications in Section~\ref{sec4}. Throughout the paper, we adopt a cosmology with $H_0=70$ km s$^{-1}$ Mpc$^{-1}$, $\Omega_{\rm m}=0.3$, and $\Omega_{\Lambda}=0.7$.

\begin{figure*}[t]
\centering
\includegraphics[width=\textwidth]{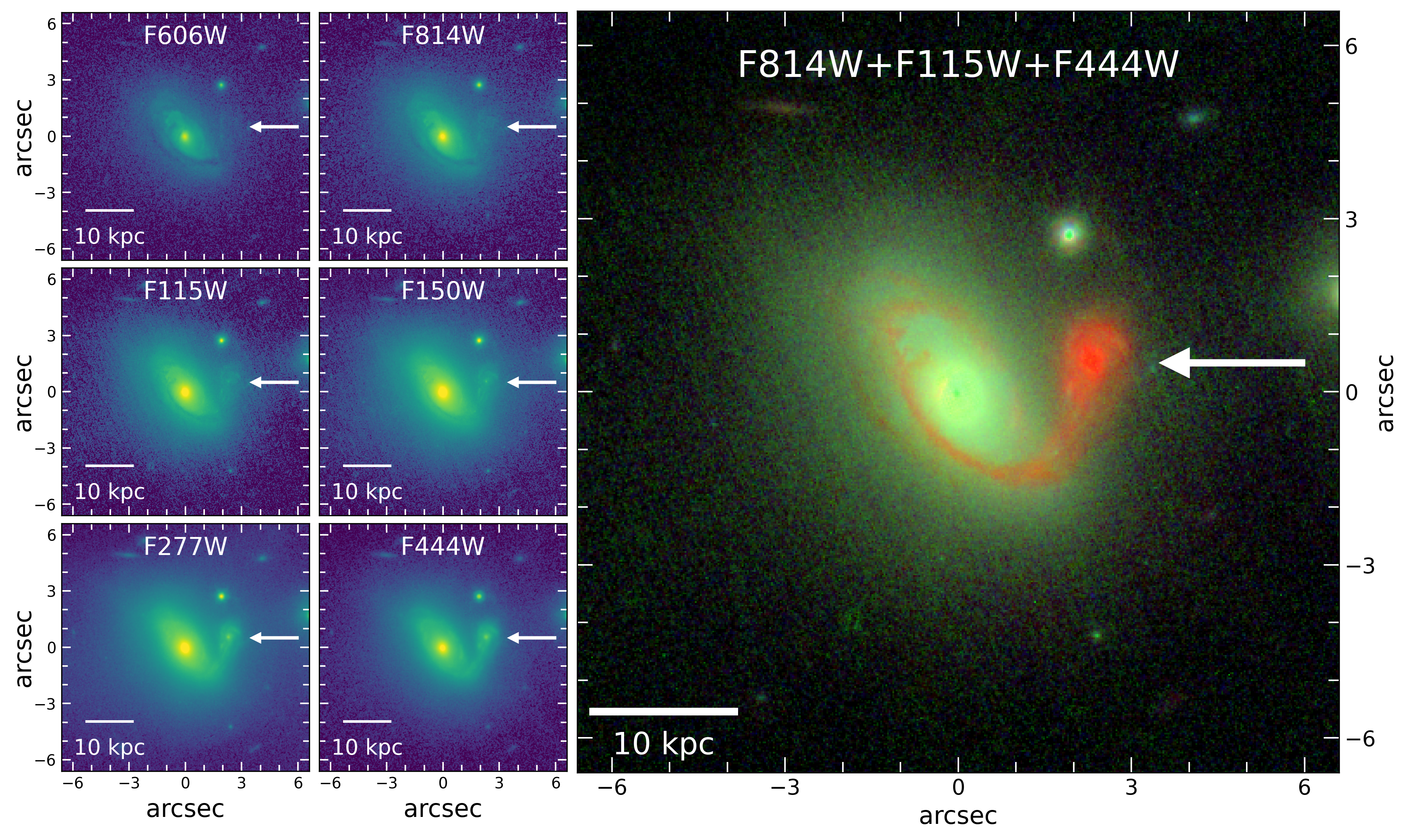}
\caption{Single-band images (left) and pseudo-color image (right) of the CW-BGG-1 system. Label in each panel indicates the name of the band: HST ACS F606W and F814W in the top row; JWST NIRCam F115W, F150W, F277W, and F444W in the middle and bottom rows. The pseudo-color image is generated using F814W (blue), F115W (green), and F444W (red). The white arrow in each panel indicates the location of the satellite CW-BGG-1-C. The bottom-left scale bar indicates 10 kpc. North is up and east is left.}\label{fig1}
\end{figure*}

\section{Data and Methods}\label{sec2}

\subsection{Imaging Data}
The system CW-BGG-1 (RA=149.919087$^{\circ}$, DEC=2.600944$^{\circ}$) was discovered from the publicly available James Webb Space Telescope (JWST) NIRCam imaging data from the COSMOS-Web project \citep{COSMOS-Web}. COSMOS-Web has NIRCam imaging in four filters: F115W, F150W, F277W, and F444W. To obtain optical coverage, we make use of archival Hubble Space Telescope (HST) ACS F814W mosaic (v2.0; pixel scale of 0.03 arcsec/pixel) from the COSMOS HST/ACS survey \citep{Koekemoer+2007ApJS, Massey+2010MNRAS} and archival HST/ACS F606W exposures (PID: 13657; PI Jeyhan Kartaltepe). We obtain calibrated, CTE-corrected ACS F606W exposures from MAST (\url{https://mast.stsci.edu/}) and use \texttt{drizzlepac} \citep{2012drzp.book.....G} to align astrometry, mask cosmic rays, and produce the final mosaic. We reduce the raw NIRCam data and produce the final mosaics using \texttt{jwst} calibration pipeline v1.15.1 \citep{Bushouse_JWST_Calibration_Pipeline_2024} and custom scripts as described in \citep{2024ApJ...962...93Z, NEXUS-EDR}. We select nearby, bright, isolated stars to construct point spread function (PSF) models using \texttt{PSFEx} \citep{Bertin2011ASPC}, following the same procedures described in \citep{2024ApJ...962...93Z}. Final NIRCam and ACS mosaics have the same pixel scale of 0.03 arcsec/pixel. We also generate PSF-matching kernels to match the PSFs in the F606W, F814W, F115W, F150W, and F277W filters to those in the F444W filter using \texttt{pypher} \citep{pypher}. Figure~\ref{fig1} shows the monochromatic images in all six filters and the composite RGB (F444W-F115W-F814W) image of the CW-BGG-1 system.

\begin{figure*}[t]
\centering
\includegraphics[width=0.8\textwidth]{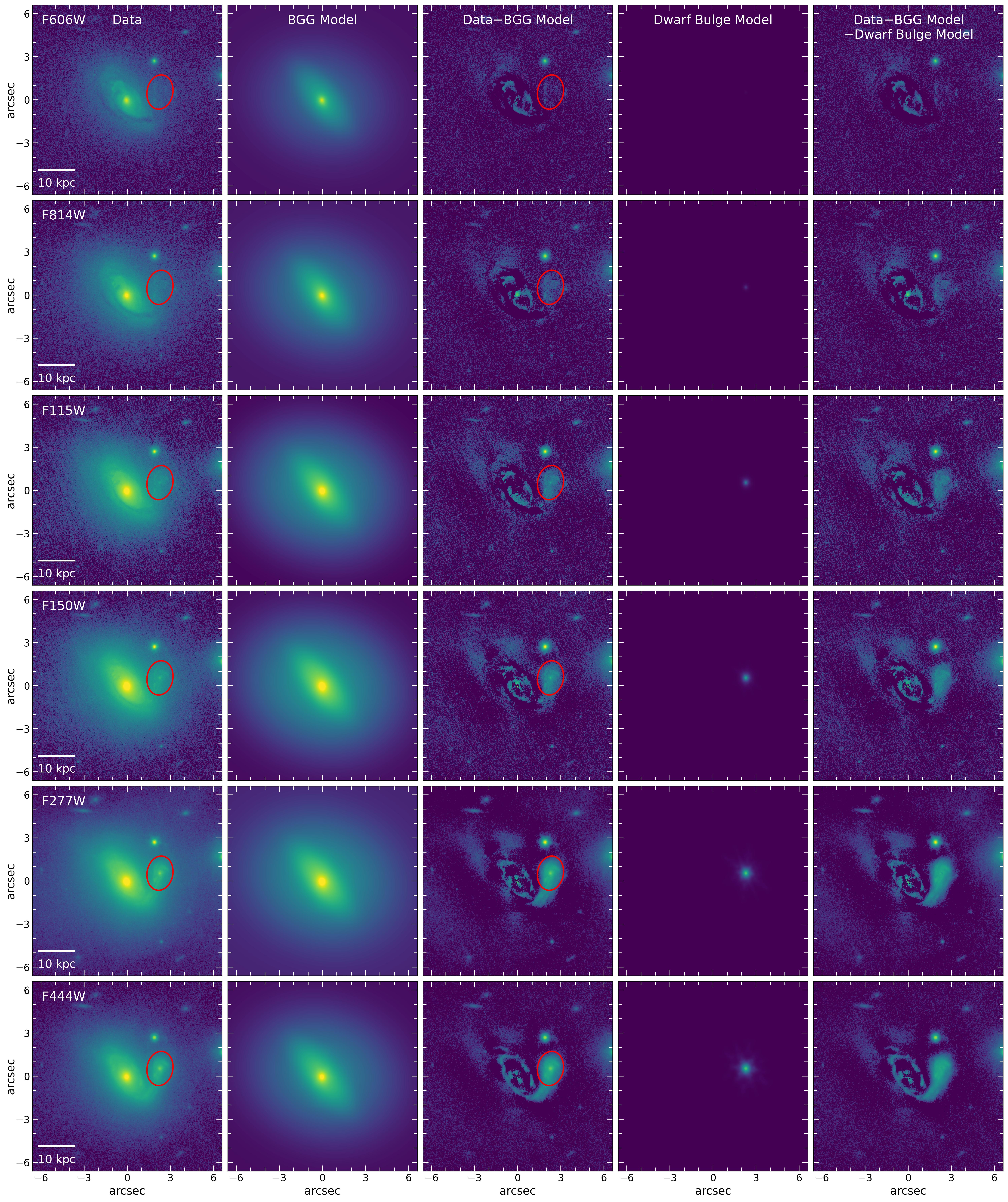}
\caption{Two-dimensional modeling of CW-BGG-1 and CW-BGG-1-C. Columns from left to right show the data, best-fit BGG model, data$-$BGG model, bulge model of the CW-BGG-1-C, and data$-$BGG model $-$ CW-BGG-1-C bulge model in six HST/ACS and JWST/NIRCam bands. BGG model is constructed by fitting four \sersic components to CW-BGG-1 after masking dust lanes, CW-BGG-1-C and its tidal tails, and neighboring sources using \texttt{GALFITM}. CW-BGG-1-C is modeled using three \sersic\ components (one for the bulge, one for the disk, and one for the extended envelope) in data$-$BGG model image after masking tidal tails. Red ellipses denote the aperture used to measure the flux of CW-BGG-1-C.}\label{fig2}
\end{figure*}

\subsection{Photometry}\label{sec2.2}
For the global photometry, which includes fluxes from both CW-BGG-1 and CW-BGG-1-C, we adopt measurements from the COSMOS2020 catalog \citep{2022ApJS..258...11W}, covering ultraviolet (UV) to mid-infrared (MIR), and the COSMOS ``Super-deblended'' photometry catalog \citep{2018ApJ...864...56J}, covering MIR to far-infrared (FIR). We use \texttt{SExtractor} AUTO fluxes in the CFHT/MegaCam $u$ band, Subaru/Suprime-Cam $grizy$ bands, VISTA/VIRCAM $YJHK_{S}$ bands, Spitzer/IRAC 3.6 $\mu$m and 4.5 $\mu$m bands from COSMOS2020 catalog. We use flux in Spitzer/IRAC 5.8 $\mu$m and 8 $\mu$m bands, Spitzer/MIPS 24 $\mu$m band, Herschel/PACS 100 $\mu$m and 160 $\mu$m bands, and Herschel/SPIRE 250 $\mu$m, 350 $\mu$m, and 500 $\mu$m bands from the COSMOS ``Super-deblended'' photometry catalog.

To extract photometry for CW-BGG-1-C, we first model and subtract the emission from CW-BGG-1 in six ACS and NIRCam bands simultaneously using \texttt{GALFITM} \citep{Haussler+2013MNRAS}. The emission from CW-BGG-1 is fitted with four \sersic\ components after masking dust lanes, CW-BGG-1-C, and neighboring sources using F606W$-$F444W and F444W images. Following the procedures described in \citep{Haussler+2022A&A}, we hold the effective radius ($R_e$), \sersic\ index ($n$), axis ratio, and position angle of each \sersic\ component constant across six bands (wavelength-independent) and allow position and magnitude free to vary in different bands. 
We note that although color gradient exists within subcomponents of individual galaxies, leading to wavelength-dependent variations in $R_e$ and $n$, these effects are generally subtle compared to the uncertainties in measuring the galaxy as a whole in most cases \citep{Haussler+2022A&A}.
The fitting results are presented in Figure~\ref{fig2}. After subtracting CW-BGG-1's emission, we perform aperture photometry in the PSF-convolved (to F444W) images using an elliptical aperture with a semi-major axis of 1.05 arcsec, a semi-minor axis of 0.75 arcsec, and a position angle of 80 degree with respect to the right counter-clockwise using \texttt{photutils} \citep{photutils}. We note that we do not include all the fluxes from tidal tails and dust lanes, since their impact on the stellar mass is minimal. Tidal tails contribute $\lesssim10\%$ in the F115W and F150W bands (restframe $\sim1$ $\mu$m). Tidal tails appear as absorption features (negative flux) in the F606W filter, while from F606W to F150W filters for dust lanes. To refine the background error budget of aperture fluxes, we measure the root-mean-square of fluxes within empty, randomly placed circular apertures with the same area as that of the companion, following methods in previous works \citep{2005ApJ...624L..81L, 2011ApJ...735...86W, NEXUS-EDR}.

To obtain the structure and flux of the CW-BGG-1-C bulge, we model its emission with three \sersic\ components using \texttt{GALFITM}: one for the bulge, one for the disk, and one for the extended envelope. Tidal tails are masked during the fitting. $R_e$, $n$, axis ratio, and position angle are held constant across six bands. 

Finally, a $10\%$ uncertainty is added in quadrature to all the fluxes to account for systematic uncertainties across different imaging surveys for spectral energy distribution (SED) modeling.

\begin{table*}[]
\caption{{\bf Model parameters of FSPS and the derived physical parameters}}
\centering
\begin{tabular}{l l l}
\hline
\hline
Parameters & Physical meaning & Priors \\\hline
\multicolumn{3}{c}{FSPS parameters} \\\hline
$M_\mathrm{total}$ ($M_\odot$)  & Total mass formed in the star formation & LogUniform($10^5$, $10^{13}$) \\
$t_\mathrm{age}$ (Gyr) & Age of the galaxy & Uniform(0, 13.8) \\
$Z$ ($Z_\mathrm{\odot}$) & Stellar metallicity & Uniform($-2$, 0.19) \\
$\tau_V$ & Optical depth & Uniform(0, 10) \\
$\tau_\mathrm{SFH}$ & SFH decay time scale & Uniform(0, 10) \\
$\gamma_\mathrm{DL07}$ & Fraction of dust in photon-dominated region region & Uniform(0, 0.2) \\
$U_\mathrm{min}$ & Minimum interstellar radiation field intensity & Uniform(0.1, 25)\\
$q_\mathrm{PAH}$ ($\%$) & Fraction of dust mass in polycyclic aromatic hydrocarbons & Uniform(0.5, 7) \\
\hline
\multicolumn{3}{c}{Derived physical parameters} \\\hline
$M_*$ ($M_\odot$) & Current stellar mass & \\
$A_V$ (mag) & $V$-band attenuation, $A_V = 2.5\log(e)\,\tau_V $ & \\
$L_\mathrm{IR}$ ($\mathrm{erg\,s^{-1}}$) & Infrared luminosity integrated over 8--1000 $\mu$m & \\
$M_\mathrm{dust}$ ($M_\odot$) & The cold dust mass derived from the dust emission model \\\hline
\end{tabular}\\
{\textbf{Note}: The left column lists parameter names. The second column explains the physical meaning of each parameter. The third column shows the priors used in the SED fitting.}
\label{tab:sed}
\end{table*}

\subsection{SED Modeling}

We conduct a joint fit to the total ultraviolet (UV)-to-far infrared (FIR) SED and the spatially-decomposed CW-BGG-1-C SED to constrain the physical properties of CW-BGG-1 and its companion simultaneously.  We incorporate the Flexible Stellar Population Synthesis \citep[\texttt{FSPS};][]{2010ascl.soft10043C} model through its \texttt{Python} wrapper ({\url{dfm.io/python-fsps}) to model the galaxy SED with the MIST \citep{2016ApJS..222....8D,2016ApJ...823..102C} isochrone model, MILES spectra library \citep{2006MNRAS.371..703S,2011A&A...532A..95F}, and the \citet[DL07;][]{2007ApJ...657..810D} dust emission templates.  The FSPS model assumes that the stellar emission attenuated by the dust is re-emitted in the infrared. We adopt the \citet{2003PASP..115..763C} initial mass function and a delayed exponential decay star formation history model.  After the star formation starts at a certain time, the instantaneous SFR of the galaxy is proportional to $t e^{-t/\tau}$, where $\tau$ is the decay time scale. In practice, we choose $\texttt{sfh}=4$ in FSPS and fix the starburst fraction to be zero ($\texttt{fburst}=0$).  We adopted the \citet{2000ApJ...533..682C} dust attenuation. The FSPS parameters involved in the modeling of a galaxy SED and the derived physical parameters are listed in Extended Table~\ref{tab:sed}.  

We emphasize that a delayed-$\tau$ SFH model and the Calzetti attenuation model are preferred in this work for simplicity.  We have tested to include more complicated SFH, for example, adding a recent starburst, and two-component dust attenuation model \citep[e.g.,][]{2013ApJ...775L..16K}.  However, we did not find the more complex models affecting any of our main conclusions.  We incorporate the same models for CW-BGG-1 and its companion, except that we add the nebular emission for CW-BGG-1-C model mainly to account for the potential contribution of the strong optical emission lines (e.g. H$\alpha$ and H$\beta$).  Since the metallicity and ionization parameters are difficult to be constrained by the broad-band SED fitting, we tie the gas-phase metallicity to the stellar metallicity \citep{2021ApJS..254...22J} and fix ionization parameter to be 0.01.  We note that the nebular emission lines do not influence our fitting results regardless of the detailed treatments of the gas phase metallicity and the ionization parameter.

We incorporate the \texttt{dynesty} \citep{2020MNRAS.493.3132S} code to conduct the Bayesian posterior inference.  We used the dynamic nested sampling and the random walk method with 1000 initial sampling points.  The sampling stops when the logarithmic evidence increase is less than 0.05.  This method guarantees that the sampling converges to the global optimal parameter region and that the posterior estimate of the model parameters is accurate.

We also model the bulge SED of CW-BGG-1-C measured from the central \sersic\ model from \texttt{GALFITM} decomposition. The SED fitting incorporated the same models and priors as the joint fitting to CW-BGG-1-C. We ignored the parameters of the dust emission model because there is no data to constrain them meaningfully.

\subsection{TNG Simulation}

We use the cosmological magneto-hydrodynamical simulation IllustrisTNG project \cite[hereafter TNG;][]{Marinacci2018, Naiman2018, Nelson2018, Pillepich2018a, Springel2018} to study the BGG of galaxy groups with similar properties as our system. These simulations are run using the moving-mesh code \textsc{AREPO} \citep{Springel2010} and include realistic prescriptions for baryonic physics, such as star formation, stellar evolution, gas heating, gas cooling, growth of black holes, and AGN feedback \citep{Weinberger2017,Pillepich2018b}. There are three simulation boxes: TNG300, TNG100, and TNG50, with progressively increasing mass resolution and decreasing volume. Each run contains 100 snapshots, approximately equally spaced in redshift between $z \sim 20$ and $z = 0$. Among the three boxes, we consider the TNG50 simulation \citep{Nelson2018, Pillepich2018a}, which spans $35\,\text{Mpc}\cdot h^{-1}$ along each side, as ideal for studying dwarf galaxies in a $z = 0$ universe. This is because it has a baryonic mass resolution of $m_{\text{baryon}} \sim 8.5 \times 10^{4}\, M_{\odot}$ and a dark matter resolution of $m_{\text{DM}} \sim 4.5 \times 10^{5}\, M_{\odot}$, which are necessary to resolve galaxies down to $\gtrsim 100$ star particles. It provides an average spatial resolution of star-forming ISM gas of $\sim 100$–200 pc. We choose TNG50 over TNG100 and TNG300 because we prioritize resolving dwarfs with stellar masses $\log(M_{*}/M_{\odot}) = 9$ over a large simulation volume. The simulated galaxies in TNG50 reside in subhalos identified using the SUBFIND algorithm \citep{Springel2001}. This algorithm is based on the friend-of-friend (FoF) prescription, which uses a linking length of $b = 0.2$ times the mean inter-particle distance. It assigns subhalos in a hierarchical manner, with all subhalos assigned to a FoF group belonging to the parent dark matter halo, which we shall refer to as the host halo henceforth. 

We select galaxy groups from the TNG50 simulation at \(z=0.2\) with similar halo and stellar masses as our target, with halo masses in the range \(10^{12.5} \leq M_{\text{halo}}/M_\odot \leq 10^{14}\) and central galaxy stellar masses between \(10^{11}\) and \(10^{12}\,M_\odot\). Within these groups, we identify dwarf satellite galaxies with stellar masses in the range \(1 \times 10^{9} \leq M_* / M_\odot < 1 \times 10^{10}\). For each satellite, we obtain their stellar mass, total cold gas mass, and orbital velocity, with stellar mass and total cold gas mass calculated within twice the $R_e$ of the stellar mass. In the TNG50 simulation, cold gas refers to gas that is assumed to have a temperature below $10^4$ K. For our analysis, we take the ``star-forming gas" (where $\text{SFR} > 0\, M_{\odot}\, {\rm yr}^{-1}$) as cold, as this gas is, by mass, assumed to be at temperatures $\ll 10^4$ K. This is most analogous to many common observational ``cold" gas measures, such as neutral or molecular gas. The rest of the gas in the simulation, with a temperature above this threshold, is considered hot gas.

To investigate the orbital dynamics of satellite galaxies within their host halos, we reconstruct their three-dimensional trajectories across snapshots at $z \geq 0.2$. By analyzing the satellite-host distance as a function of redshift, we identify the pericenter as the point of closest approach (a local minimum at $0.2 \leq z \leq 0.5$) and the nearest preceding apocenter as the point of maximum separation (local maximum at $z > 0.2$). At these orbital points, we extract the satellite's distances to the BGG at both the pericenter ($r_{\text{peri}}$) and apocenter ($r_{\text{apo}}$). The elongation of the orbit is estimated using orbital ellipticity ($e$) defined as 
\[
e = 1 - \frac{r_{\text{peri}}}{r_{\text{apo}}},
\]
where $r_{\text{peri}}$, $r_{\text{apo}}$ indicate the separation between the BGG and satellite at pericenter and apocenter, respectively. Due to the limited number of saved snapshots in the simulation, the real pericenters of these satellite galaxies are somewhat smaller than $r_{\text{peri}}$ calculated here. If the separation of a satellite's orbit increases monotonically at $0.2 \leq z \leq 0.5$, we extend this redshift range to select the pericenter for orbital ellipticity estimation only. 

Our final sample consists of 270 satellites across 44 galaxy groups, all located within the virial radius ($r_{\rm{peri}}<r_{\rm{vir}}$). We exclude three satellites without apocenter measurements, as they are departing from the BGG after their formation. We define first close-separation pericentric passage if satellites move inside $0.1\times r_{\rm vir}$ for the first time at $0.2 \leq z \leq 0.5$.

TNG50 assumes a flat $\Lambda$CDM cosmology with parameters adopted from \citep{Planck2016}: $\Omega_{\Lambda} = 0.6911$, $\Omega_{\mathrm{m}} = 0.3089$, $\Omega_{\mathrm{b}} = 0.0486$, $\sigma_{8} = 0.8159$, $n_{s} = 0.9667$, and $h = 0.6774$. We ignore the tiny difference on mass ($\ll$0.1 dex) due to the different cosmologies adopted in TNG50 and this paper. 

\begin{figure*}[t]
\centering
\includegraphics[width=0.55\textwidth]{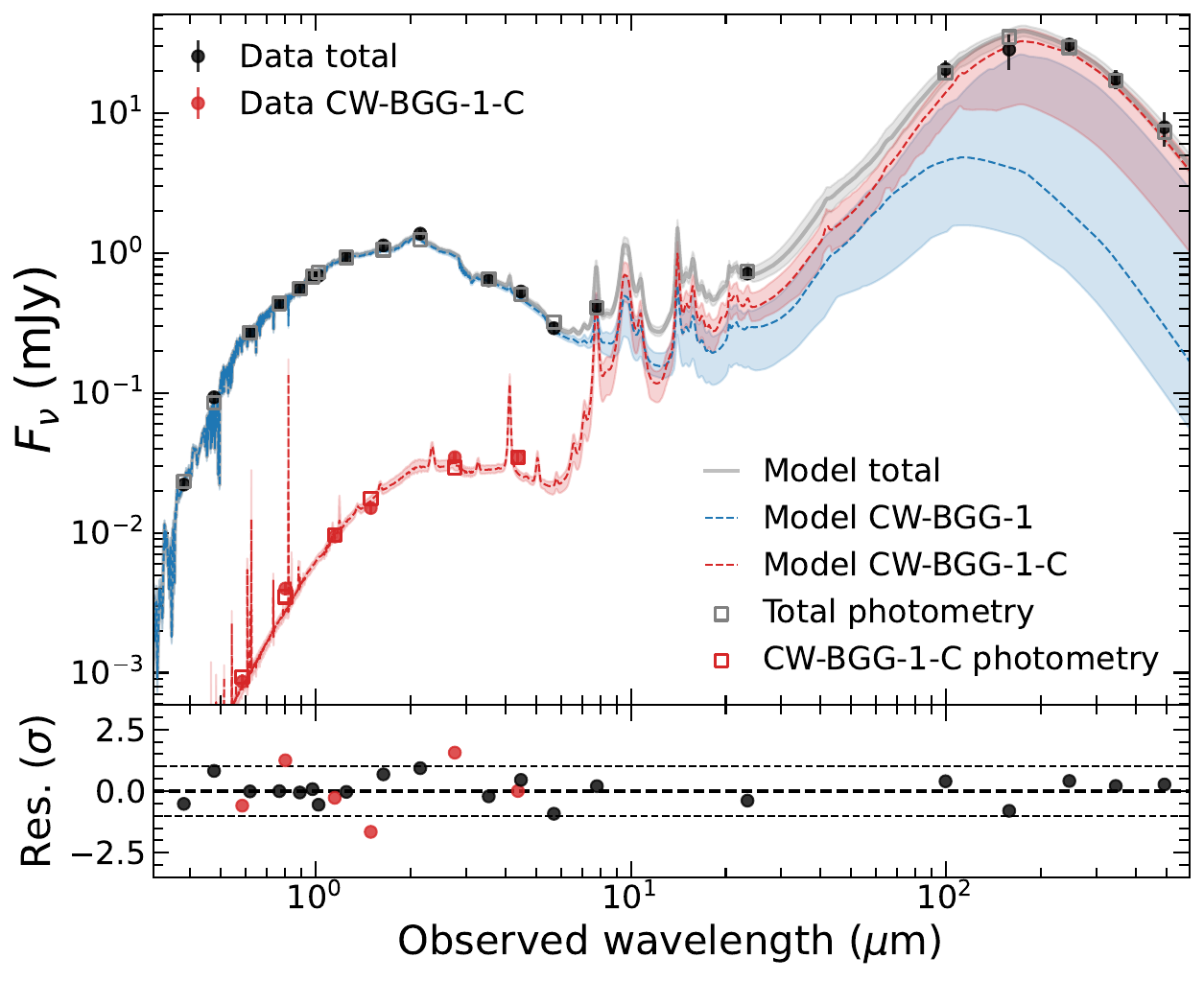}
\includegraphics[width=0.44\linewidth]{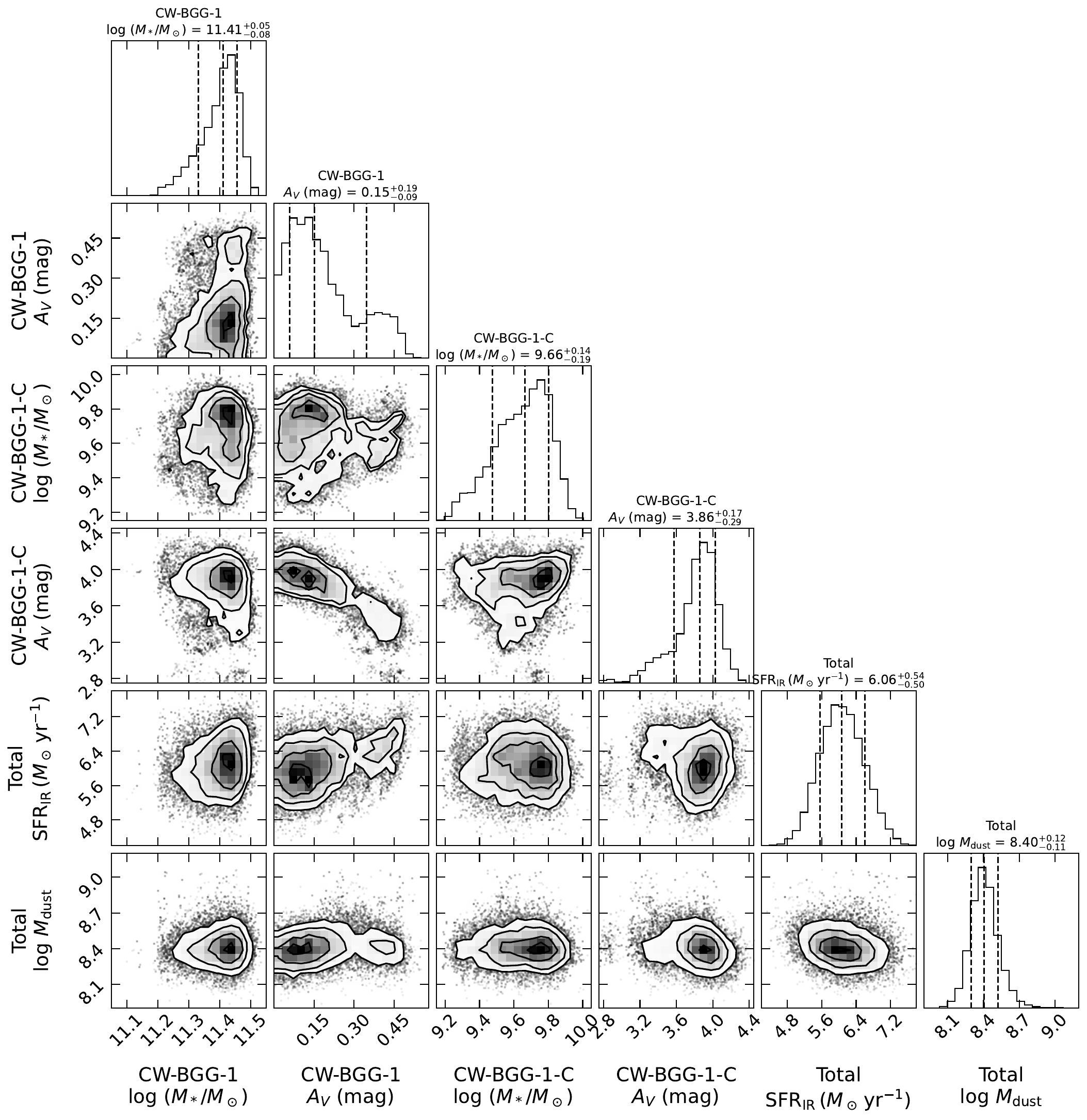}
\caption{Joint spectral energy distribution modeling of the CW-BGG-1 system (left) and posterior distributions of the derived physical parameters (right). The black dots represent the total SED of the CW-BGG-1 system while the red dots represent SED of CW-BGG-1-C after decomposing emission from CW-BGG-1. The empty gray and red squares are the model photometric fluxes at the corresponding observed bands of the data for CW-BGG-1 and CW-BGG-1-C, respectively. The gray, blue, and red curves are the median total, median CW-BGG-1, and median CW-BGG-1-C model SEDs from our Bayesian nested sampling posteriors, respectively. The shaded regions indicate the 16th-to-84th percentile confidence ranges. The lower panel displays the residual data with respect to the median models in the unit of observed uncertainties. Most of the data points are within 1$\sigma$ deviations.  The right panel displays the posterior distribution of the physical parameters derived from the joint SED fitting.  The first four parameters are the stellar mass and $V$-band extinction ($A_V$) of the two galaxies, respectively. The last two parameters are the total star formation rate (SFR) based on the total infrared (integrated over 8-1000~$\mu$m) luminosity and total dust mass of the CW-BGG-1 system.}
\label{fig3}
\end{figure*}

\begin{figure}
\centering
\includegraphics[width=0.45\textwidth]{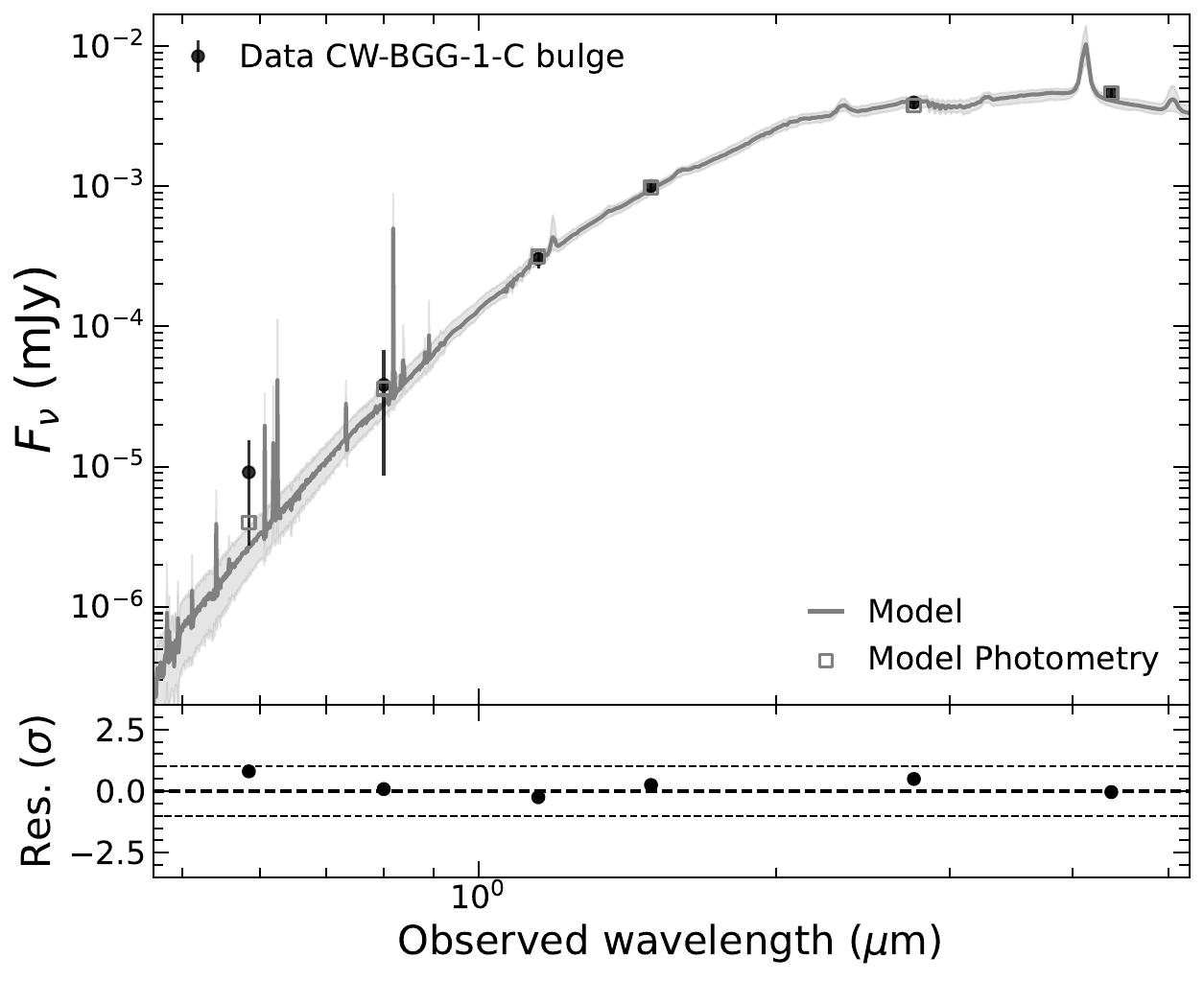}
\includegraphics[width=0.45\textwidth]{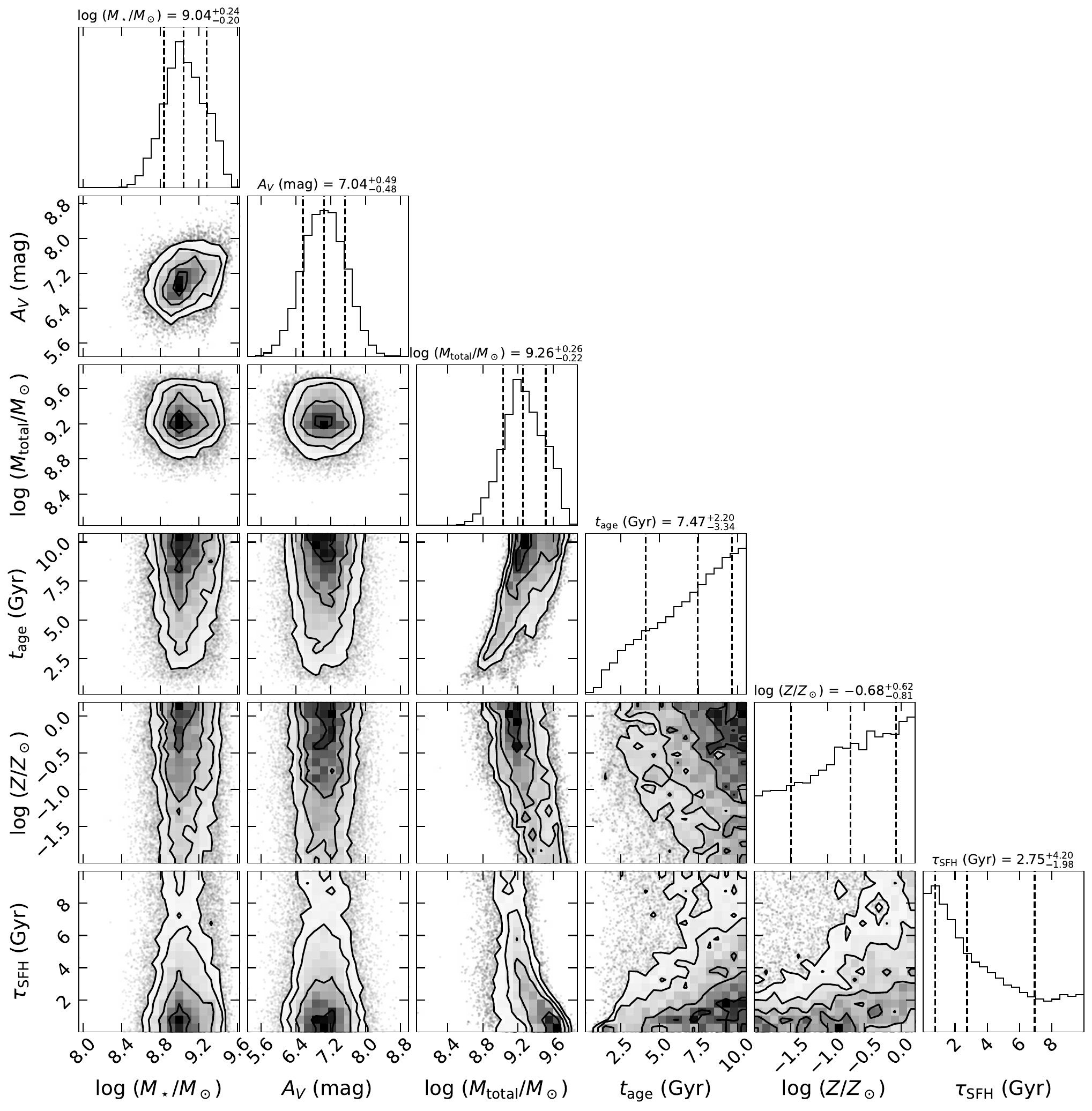}
\caption{SED data (top) and posterior distributions of model fitting parameters (bottom) to the bulge of CW-BGG-1-C. The upper panel displays the SED data (black points) and the model median (gray curve) and 16-to-84 confidence interval (gray shaded region) of the fitting to CW-BGG-1-C bulge SED, with fitting residual showing underneath. The bottom panel displays the posterior distributions of model fitting parameters and the derived physical parameters.}
\label{fig4}
\end{figure}

\begin{figure}[t]
\centering
\includegraphics[width=0.48\textwidth]{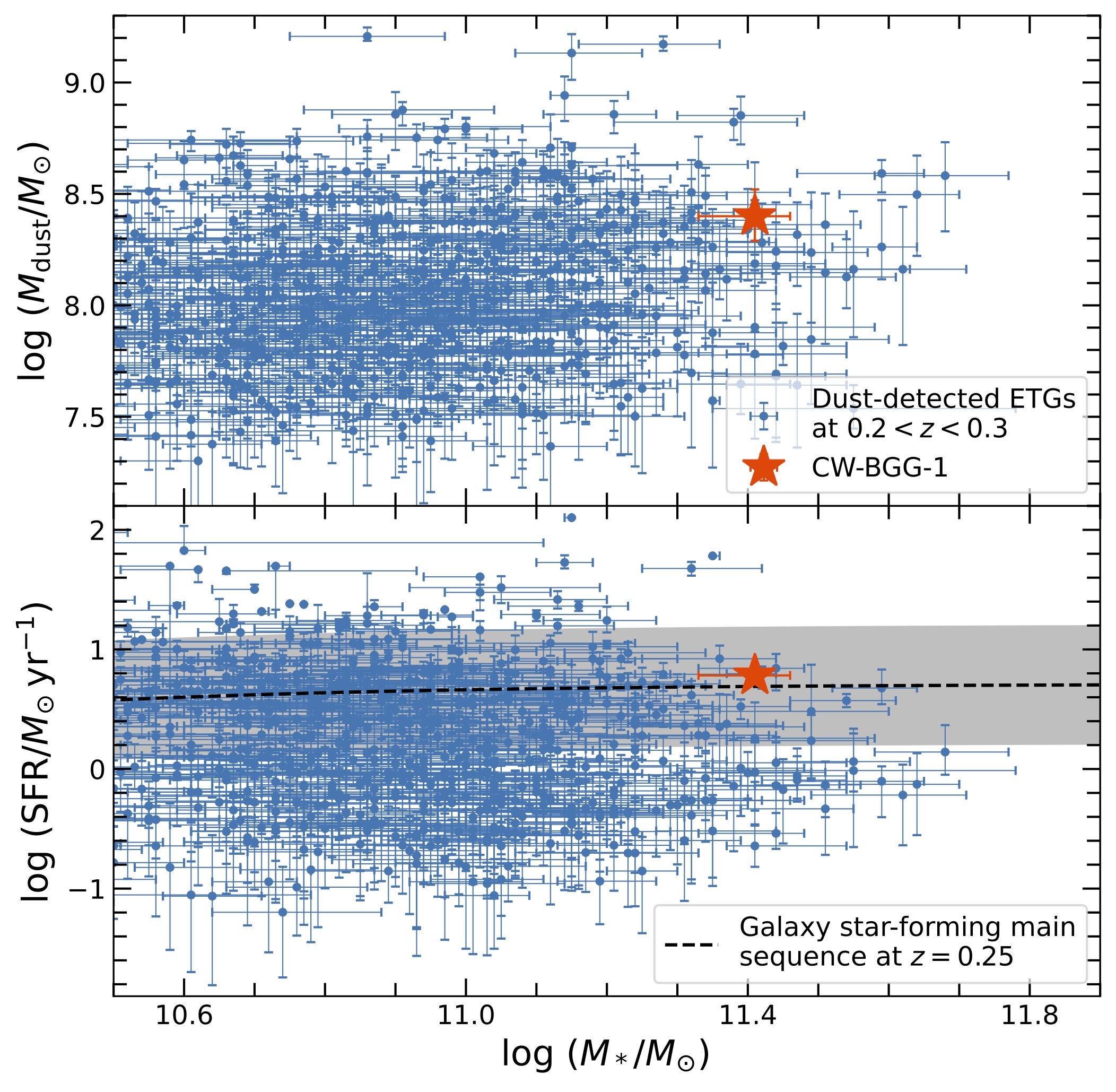}
\caption{Dust mass ($M_{\rm dust}$) and star formation rate (SFR) versus stellar mass ($M_*$). The red star denotes the CW-BGG-1 system while the blue dots represent dust-detected early-type galaxies at $0.2<z<0.3$ from \citep{2023ApJ...953...27L}. Error bars indicate 1$\sigma$ uncertainties. The black dashed curve and gray shaded stripe denotes the galaxy star-forming main sequence at $z=0.25$ taken from equation 14 in \citep{SFMS} and its $\pm0.5$ dex dispersion, respectively. CW-BGG-1 is located on the galaxy star-forming main sequence and is among the dustiest ETGs at $0.2<z<0.3$ in the massive end ($M_*>10^{10.5}\,M_{\odot}$).}\label{fig5}
\end{figure}

\begin{figure*}[t]
\centering
\includegraphics[width=0.8\textwidth]{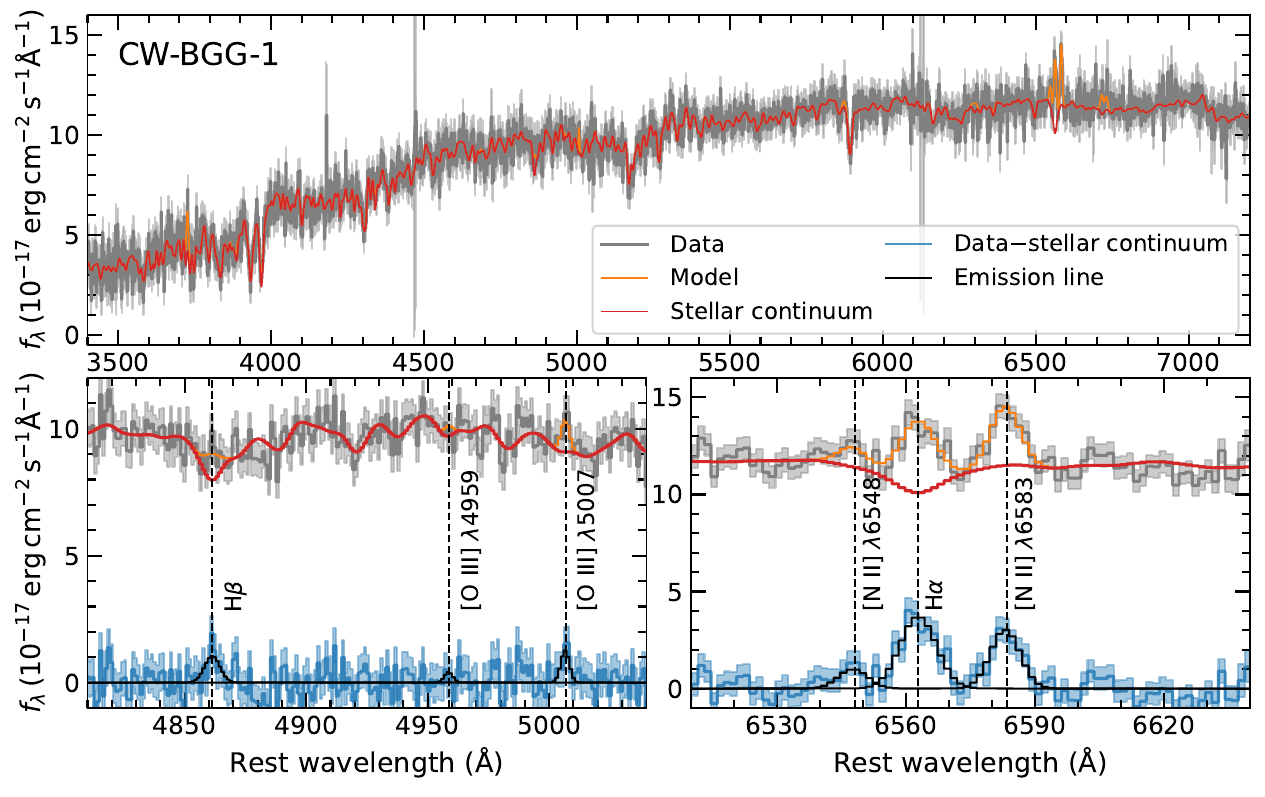}
\includegraphics[width=0.8\textwidth]{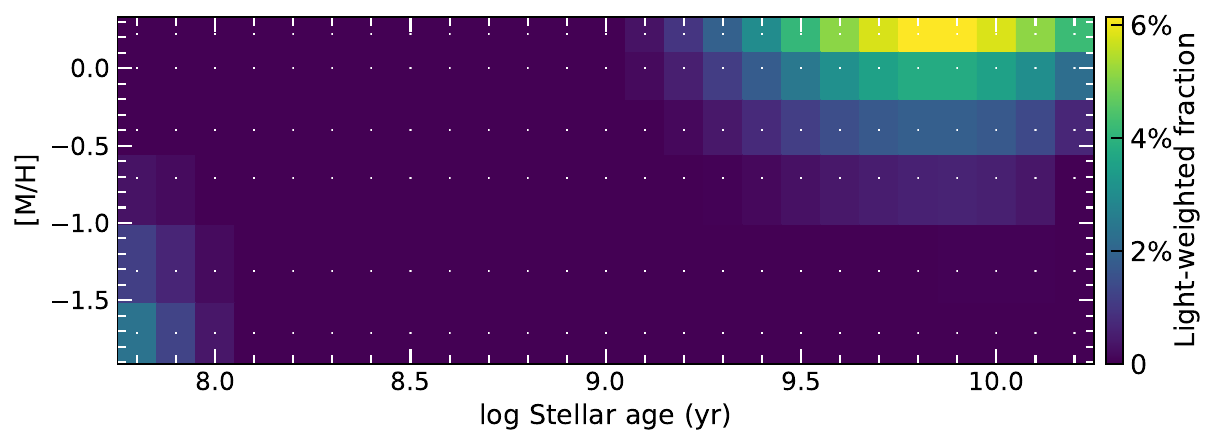}
\caption{Full spectral modeling to CW-BGG-1. Gray curve and shaded region represent the restframe spectrum of CW-BGG-1 and its uncertainty. Orange curve represents combined model from stellar continuum (red curve) and emission lines (black curve) using \texttt{pPXF} \citep{pPXF}. Blue curve represents emission line-only spectrum after subtracting the continuum (data$-$stellar continuum). Top panel shows the full spectrum fitting result, while middle panels show zoom-in views around \Hb\ and \Ha, respectively. Prominent emission lines are labeled and indicated with vertical dashed lines. Bottom panel shows the stellar population distribution in metallicity ([M/H]) and stellar age plane color-coded by light-weighted fraction.}\label{fig6}
\end{figure*}

\begin{figure*}[t]
\centering
\includegraphics[width=0.8\textwidth]{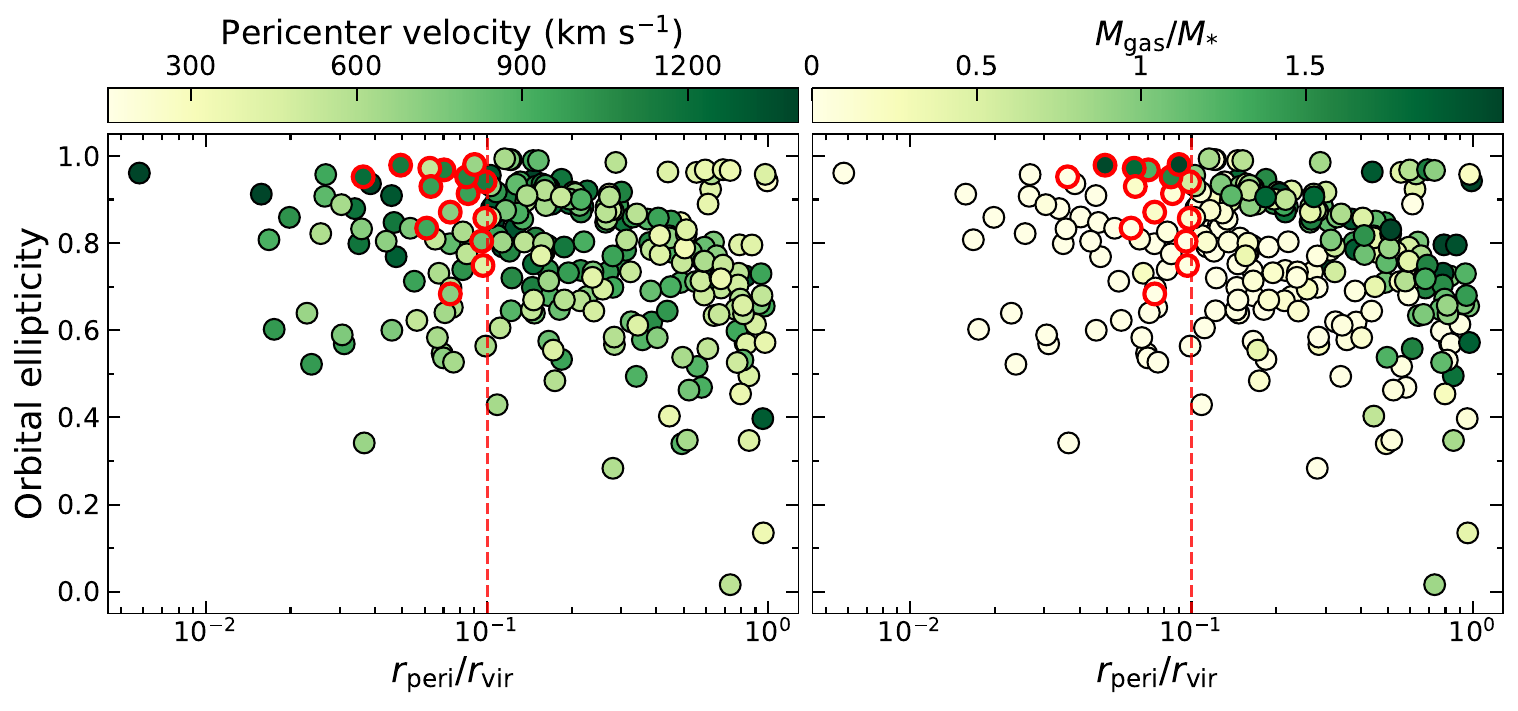}
\includegraphics[width=0.8\linewidth]{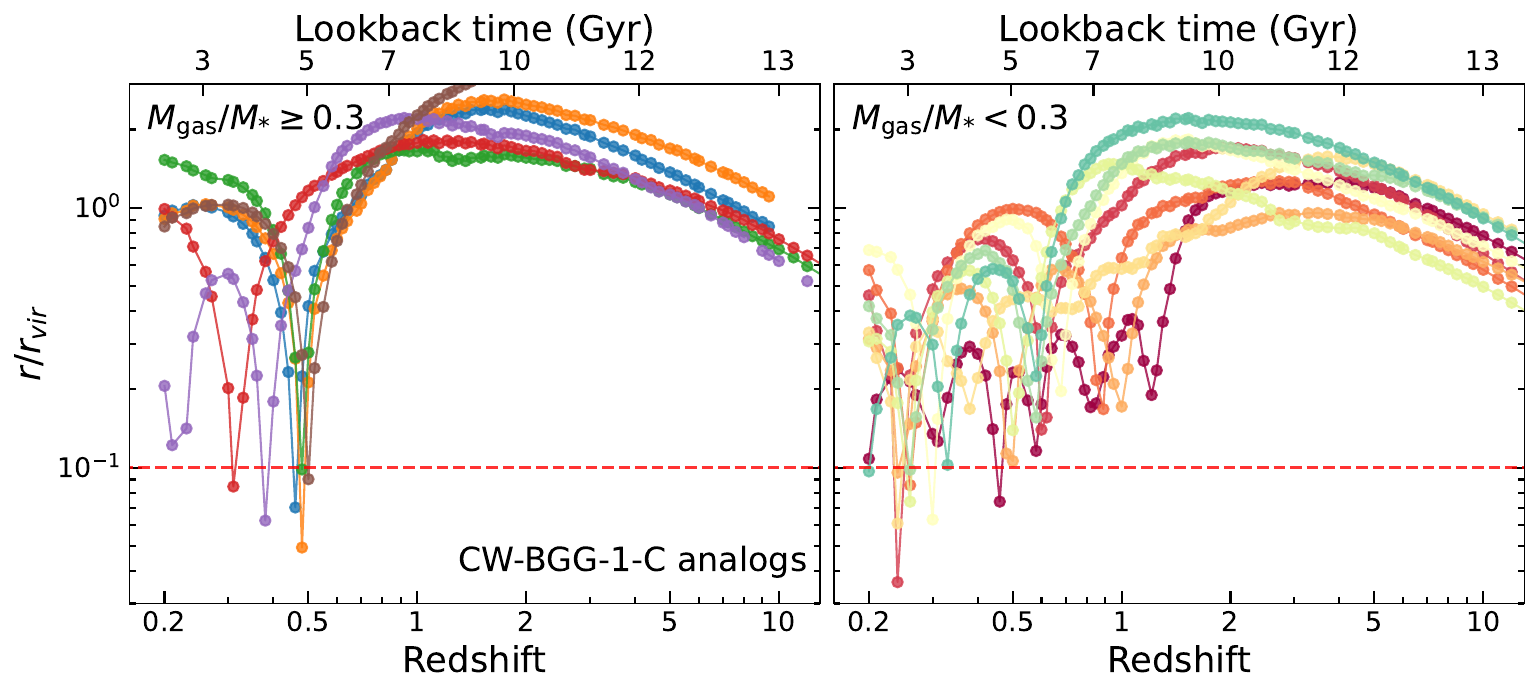}
\caption{Orbital properties of satellites in galaxy groups from the TNG50 simulation \citep{Nelson2018, Pillepich2018a}. {\bf Top:} Orbital ellipticity versus pericenter distance ($r_{\rm peri}$) normalized by the virial radius of the group ($r_{\rm vir}$). Satellites undergoing their first close-separation pericentric passage at $0.2\leq z \leq 0.5$ are marked with red circles, where close-separation is defined as $r_{\rm peri}/r_{\rm vir}\leq 0.1$ (red vertical dashed line). The left panel is color-coded by satellites' velocity at pericenter, while the right panel is color-coded by their total cold gas fraction ($M_{\rm gas}/M_*$) measured at the first preceding apocenter. {\bf Bottom:} Orbital histories showing satellites' distance to BGG ($r$, normalized by $r_{\rm vir}$) as a function of redshift for satellites undergoing the first close-separation pericentric passage at $0.2\leq z \leq 0.5$ (objects with red circles in the {\bf top} panel). The left panel displays gas-rich ($M_{\rm gas}/M_*\geq 0.3$) satellites (CW-BGG-1-C analogs), while the right panel shows gas-poor satellites.}\label{fig7}
\end{figure*}

\section{Results}\label{sec3}
\subsection{System Discovery and Basic Properties}

We present the first direct evidence of cold gas transport through a gas-rich minor merger in CW-BGG-1, a massive lenticular galaxy at $z=0.2475$ in the COSMOS field. It is the BGG of a massive galaxy group with a halo mass ($M_{\rm halo}$) of $\sim 10^{13.4} \, M_{\odot}$ \citep{2012ApJ...753..121K, 2019MNRAS.483.3545G, 2020MNRAS.499...89W}, with a prominent galactic-scale ($\sim 10$ kpc) dust lanes (Figure~\ref{fig1}) and a stellar mass ($M_*$) of $10^{11.4 \pm 0.1} \, M_{\odot}$. A close companion, CW-BGG-1-C, located at a projected separation of $\sim9.3$ kpc, is barely visible in the optical but brightens at longer wavelengths, indicating substantial dust obscuration. The physical connection between CW-BGG-1 and CW-BGG-1-C is evidenced by the presence of stripped dust and gas from the satellite, which directly link dust lanes aligned with the plane of the BGG’s stellar disk. 

Joint SED modeling (Figure~\ref{fig3}) of the two galaxies shows that CW-BGG-1-C is a low-mass, dusty dwarf galaxy with a stellar mass of $10^{9.7\pm0.1} \, M_{\odot}$, a very high global visual extinction ($A_V=3.9\pm0.2$ mag), and a stellar mass ratio of $\sim1:56$ relative to CW-BGG-1. Its pronounced red optical-to-IR colors and elevated fluxes in F277W and F444W are driven by a combination of substantial dust extinction and a strong hot dust continuum that includes the polycyclic aromatic hydrocarbon (PAH) 3.3 \micron\ feature.

\subsection{Morphology of the CW-BGG-1 System}

Using the detailed structural modeling described in Section~\ref{sec2.2}, we characterize the detailed morphology of the CW-BGG-1 system.

CW-BGG-1 is a lenticular galaxy composed of a compact bulge, an inner lens, an elongated stellar disk, and an extended outer envelope. The bulge is compact, with $R_e\approx 830$ pc, $n\approx1.7$, and a bulge-to-total ratio (B/T) of $\sim18\%$ at restframe 0.9 micron (F115W). The inner lens has $R_e\approx3.3$ kpc, $n\approx1.7$, and contributes $\sim19\%$ of the total light. The disk component is elongated (axis ratio $\approx0.44$), with $R_e\approx6$ kpc, $n\approx0.5$, and a light fraction of $\sim24\%$. CW-BGG-1 is further embedded in an extended outer envelope ($R_e\approx11$ kpc) with a flat \sersic\ profile ($n\approx0.6$), likely associated with intragroup light. The relative contributions of the inner (bulge), intermediate (lens + disk), and outer (envelope) components closely resemble those of massive local ellipticals \citep{2013ApJ...766...47H}, consistent with CW-BGG-1 being a dynamically mature early-type system.

CW-BGG-1 belongs to a relatively rare population of dust-lane ETGs \citep{2012MNRAS.423...49K}, in which dust is considered good tracers of gas-rich minor mergers. These galaxies often exhibit elevated AGN activity, young stellar populations, and clear tidal features \cite[e.g.,][]{2012MNRAS.423...59S, 2015MNRAS.449.3503D}. Figure~\ref{fig1} clearly shows that CW-BGG-1-C is transporting gas and dust into CW-BGG-1 via tidal stripping, forming the distinct dust lanes \textit{in situ}. 

CW-BGG-1-C exhibits similar morphologies as late-type Sc/Sd galaxies, with a very small B/T ($\sim3\%$) at restframe 0.9 micron (F115W). Constraints from longer wavelengths data indicate that the bulge is compact, with $R_e\approx410$ pc and $n\approx1.5$. The bulge of CW-BGG-1-C is barely visible in F606W and F814W, likely due to heavy dust extinction ($A_V\approx7$ mag), and has a stellar mass of $\sim10^9\,M_\odot$ (Figure~\ref{fig4}). The stellar age of bulge tends to be old, peaking at the largest age allowed by the redshift. However, we caution that this may be due to the degeneracy between dust attenuation and stellar age, leading to overestimated stellar mass \citep{2013ARA&A..51..393C}. CW-BGG-1-C also features fragmented, loosely wounded spiral arms, which are noticeable in four NIRCam filters. 

\subsection{Interstellar Medium and Star Formation Properties of the System}

The total dust continuum emission of the CW-BGG-1 system is robustly constrained from the Herschel photometry. We derive a total dust mass of $M_{\rm dust}=10^{8.4\pm0.1} \, M_{\odot}$ by summing up the posterior distributions of $M_{\rm dust}$ of each component. We obtain a total cold gas (molecular and atomic) mass ($M_{\rm gas}$) of $10^{10.4} \, M_{\odot}$, assuming a gas-to-dust ratio (GDR) of 100, typical for local dust detected ETGs from the Herschel Reference Survey \citep{2012ApJ...748..123S}. We estimate the total star formation rate (SFR) of the system by integrating the FIR luminosity ($L_{\rm IR}$) over 8--1000 $\mu$m using the calibration from \citep{1998ARA&A..36..189K}. After correction to the Chabrier (2003) initial mass function \citep{2014ARA&A..52..415M}, we find an SFR of $6.1\pm0.6 \, M_{\odot}\, \rm{yr}^{-1}$. 

Figure~\ref{fig5} compares the $M_{\rm dust}$ and SFR of the CW-BGG-1 system with dust-detected ETGs in the Herschel/SPIRE 250 $\mu$m band at $0.2<z<0.3$ from \citep{2023ApJ...953...27L}. CW-BGG-1 is located on the galaxy star-forming main sequence \citep{SFMS} and is among the dustiest ETGs at $0.2<z<0.3$ in the massive stellar mass end ($M_*>10^{10.5}\,M_{\odot}$).

\subsection{Spectral Properties of the BGG}

CW-BGG-1 has archival optical spectrum from the Sloan Digital Sky Survey \citep{SDSS}. The spectrum was obtained with a 2 arcsec-diameter fiber covering the central $\sim7.8$ kpc of the BGG, including the bulge and part of the disk and dust lanes. We perform full spectral modeling using \texttt{pPXF} \citep{pPXF} and EMILES stellar population models \citep{EMILES}. We adopt a typical multiplicative Legendre polynomial degree of 10 to correct the continuum shape during the fit. 

Figure~\ref{fig6} shows the fitting results with zoom-in views around H$\beta$ and H$\alpha$ and stellar population distribution. We obtain a stellar velocity dispersion of 210$\pm$10 km s$^{-1}$ within the fiber after correcting the instrumental broadening. It has moderate luminous \Ha\ and \NII$\lambda\lambda$6548,6583 lines, with \Ha\ equivalent width of $\sim3.8$ \AA. All the other lines are very weak, with only \OII$\lambda\lambda$3727,3729, \Hb, \OIII$\lambda$5007, and \SII$\lambda\lambda$6716,6731 detected at a significance slightly above 3$\sigma$. Balmer decrement gives a $E(B-V)$ of 0.52 mag adopting the \citet{ccm89} extinction curve, $R_V=3.1$ and an intrinsic \Ha/\Hb=2.86. Extinction corrected \OIII$\lambda5007$ and \Ha\ luminosities are $10^{40.8}$ and $10^{41.4}$ erg s$^{-1}$, respectively. 

We find that the central region of CW-BGG-1 is dominated by old stellar populations with prominent stellar absorption features and a light-weighted age of $\sim4.7$ Gyr. Interestingly, a small fraction ($\sim6.5\%$) of the total stellar population is young (age $\lesssim 100$ Myr) and metal-poor (1--10\% solar metallicity). While the exact locations of these young stars remain uncertain, they are likely spatially associated with the stripped material or dust lanes. They could either have originated from the satellite or formed \textit{in situ} out of the stripped cold gas. Integral field spectroscopy is required to confirm the presence and spatially locate these active star forming regions.

For completeness, while the central spectrum is dominated by old stellar populations, we cannot fully exclude the presence of a weak active galactic nucleus (AGN), given the ambiguous classifications in the ``Baldwin, Phillips \& Terlevich” (BPT) diagnostic diagrams \citep{1981PASP...93....5B}. Nevertheless, the derived properties of the BGG are largely unaffected by potential weak AGN contamination.

\subsection{Close-separation Interactions in Galaxy Groups from the TNG50 Simulation}

We use the TNG50 cosmological simulation \citep{Nelson2018, Pillepich2018a} to search and characterize close-separation interactions in galaxy groups. We find that approximately $30\%$ (12/44) of massive BGGs ($M_* > 10^{11}\, M_{\odot}$) in halos of $10^{12.5-14}\,M_{\odot}$ have experienced at least one close-separation interaction with a satellite of similar stellar mass to CW-BGG-1-C ($10^{9-10}\,M_{\odot}$), with their first pericentric passage occurring at $0.2 \leq z \leq 0.5$. These interactions take place at a pericenter distance within $10\%$ of the virial radius of the group, corresponding to $\sim60$ kpc at halo mass of $10^{13.5}\, M_{\odot}$. The involved satellites account for nearly one-sixth (15/94) of the total satellite population within the virial radius of their host galaxy group. 

As expected, most of these satellites exhibit highly eccentric orbits (orbital ellipticity $>0.8$) and high velocities ($\sim800$ km/s), facilitating rapid and close encounters with their central BGGs (Figure~\ref{fig7}). Notably, 40\% (6/15) were gas-rich ($M_{\rm gas}/M_*>0.3$, with a median value of $\sim$1.5) prior to their most recent infall, making them strong analogs of CW-BGG-1-C. In contrast, the remaining nine satellites have relatively lower orbital ellipticity and pericenter velocity and were predominantly gas-poor or gas-deficit before their most recent infall. 

A detailed examination of the orbital histories of these satellite galaxies reveals that CW-BGG-1-C analogs experience their first-ever infall from beyond $0.5\times r_{\rm vir}$, while gas-poor counterparts undergo multiple crossings into $\leq 0.5\times r_{\rm vir}$ region (Figure~\ref{fig7}). The prolonged (several Gyr) environmental processing, coupled with gradual inward migration due to tidal stripping, likely leads to the cold gas depletion of these gas-poor satellites.

The close-separation gas-rich minor merger rate, averaged over 2.6 Gyr ($0.2\leq z \leq 0.5$) for the six gas-rich satellites undergoing their first infall during this time and their five host BGGs, is $\sim0.46$ Gyr$^{-1}$, in good agreement with previous estimates ($\sim0.6$ Gyr$^{-1}$) based on simple analytic toy models \citep{2019MNRAS.486.1404D}. Our results thus  suggest that gas-rich minor mergers represent a significant channel for cold gas replenishment in central group galaxies.

\section{Discussion and Conclusions}\label{sec4}

\subsection{Formation Channel of Low-mass Compact Elliptical Galaxies}

CW-BGG-1-C features a luminous bulge and an extended disk with spiral arms. If its extended structures, including the disk and outer envelope, are fully disrupted and stripped after several passes, the remaining bulge ($M_*=10^{9.0\pm0.2} \, M_{\odot}$, half-light radius of $\sim400$ pc) would become a low-mass compact elliptical galaxy, supporting the stripping mechanism being an important formation pathway for compact ellipticals in dense environment \cite[e.g.,][]{2001ApJ...557L..39B, 2018MNRAS.473.1819F}. Our SED fit to the bulge of CW-BGG-1-C (Figure~\ref{fig4}) reveals a very high dust extinction ($A_V=7.0\pm0.5$ mag). This is consistent with the quick metal-enrichment due to rapid star formation and confinement of metal-rich gas from ram-pressure and tidal force, as predicted in  \citet{2019ApJ...875...58D} and \citet{2025ApJ...979L..33B}.

\subsection{The Hidden Low-Mass, Dusty Population of Satellite Galaxies}

Satellite galaxies in galaxy groups tend to exhibit lower SFR and reduced gas content compared to field galaxies of similar mass, albeit with a bimodal distribution of SFR in low-mass objects \citep{2012MNRAS.424..232W, 2017MNRAS.466.1275B}. Low-mass, dusty, and gas-rich satellites with extreme global dust extinction ($A_V\gtrsim1$ mag) like CW-BGG-1-C have been identified before, likely due to their rarity in the local Universe, where cosmic downsizing has led to a decline in the gas content \citep{2020ARA&A..58..157T}. While beyond the local Universe, these systems are easily missed in large surveys, since their faint restframe optical emission and proximity to bright central galaxies make them extremely challenging to detect. JWST NIRCam is uniquely suited to systematically identify such satellites over a wide redshift range, thanks to its large field-of-view, broad wavelength coverage, exquisite spatial resolution, and unparalleled IR sensitivity. 

The elevated galaxy merger rate toward the earlier Universe \citep{2011ApJ...742..103L} suggests a higher probability of detecting more hidden gas-rich minor mergers or close flybys. Indeed, recent JWST observations have already uncovered a population of highly dusty dwarf galaxies at $z<2$ with a median $A_V = 3.9$ mag and a median stellar mass of $10^{7.3} \, M_{\odot}$ \citep{2023A&A...676A..76B}. This finding is unexpected given the low stellar masses of these galaxies \citep{2015ApJ...807..141P}, highlighting JWST’s unprecedented ability to characterizing such systems. As mentioned above, these galaxies are likely the progenitors of compact elliptical galaxies undergoing bursty star formation activities \citep{2019ApJ...875...58D, 2025ApJ...979L..33B}, which further emphasizes their importance in galaxy evolution.

\subsection{Relative Contribution of the ISM}

The total dust mass and SFR of the CW-BGG-1 system are robust and consistent with estimates derived from SED fitting to the global photometry under the assumption of a single galaxy. However, ETGs with dust lanes exhibit a broad range of GDR values (50--750) with an elevated median of $\sim$200 \citep{2015MNRAS.449.3503D}, suggesting that the total cold gas mass of the system may be underestimated by adopting a GDR of 100. 

Our model inference suggests that CW-BGG-1-C contributes the majority of the FIR emission (around two-thirds), consistent with the strong dust attenuation and strong MIR emission in the companion and the early-type morphology of the BGG. Stellar population modeling to both SED and optical spectrum further supports this, indicating that CW-BGG-1-C accounts for the majority of the SFR, cold dust, and gas of the system. However, we caution that the relative contribution of FIR emission remains highly degenerate between the two galaxies since we only have integrated photometric measurements. In fact, the dust-to-stellar mass ratio of the satellite is at least a factor of a few higher than the typical values of dwarf galaxies \citep[$\lesssim 1\%$; see e.g.,][]{2019A&A...626A..63D}, if two-third of the dust mass belongs to CW-BGG-1-C. While a significant fraction of dust may come from rapid star formation induced by tidal interaction \citep{2019ApJ...875...58D}, the exact amount is still highly uncertain with constraints from current data. As the main goal of this work is not to decompose the FIR emission and interstellar medium contribution, we only report the total infrared luminosity and dust mass from the SED modeling. 

To resolve the degeneracy between the two objects and fully understand the distribution of cold gas and dust, dedicated high-resolution sub-millimeter observations with interferometric telescopes, such as the Atacama Large Millimeter/submillimeter Array (ALMA) and the Northern Extended Millimetre Array (NOEMA), are necessary. These observations will map the cold gas and dust distributions and quantify the relative contributions from CW-BGG-1 and its companion.

\subsection{Concluding Remarks}

In this paper, we report the first unambiguous case of direct cold gas transport onto a BGG through an ongoing gas-rich minor merger. High-resolution JWST and HST imaging reveals a dust-rich, morphologically disturbed dwarf satellite undergoing tidal stripping at a projected separation of only $\sim9$ kpc from the massive central lenticular galaxy. The stripped material forms a continuous bridge between the two galaxies, forming the prominent $\sim$10 kpc dust lanes \textit{in situ}. Joint UV–FIR SED modeling demonstrates that the satellite is a low-mass ($\sim10^{9.7},M_{\odot}$) yet heavily dust-obscured system with global extinction $A_V\approx3.9$ mag and a stellar mass ratio of $\sim$1:56 relative to the BGG.  

The stark contrast between the overwhelmingly old stellar population of the BGG and the substantial dust and gas content of the system provides direct evidence that gas-rich minor mergers can efficiently replenish the cold ISM in massive early-type galaxies. Using the TNG50 cosmological simulation, we show that close-separation interactions involving gas-rich low-mass satellites undergoing their first infall on highly eccentric orbits occur at a rate of $\sim0.5$ Gyr$^{-1}$ in group environments, indicating that such events are not uncommon.  

Our results imply that a single gas-rich minor merger can deliver enough cold gas to partially rejuvenate a quiescent massive galaxy, potentially driving it back toward the star-forming main sequence. Combined with recent JWST discoveries of dusty dwarf galaxies at intermediate redshift, this work suggests the existence of a previously overlooked population of heavily obscured low-mass satellites in galaxy groups, which may play an important role in regulating the growth and evolutionary pathways of their central galaxies.

\begin{acknowledgments}
Support for Program number JWST-AR-03038 was provided through a grant from the STScI under NASA contract NAS5-03127. 
Jinyi Shangguan is supported by the Fundamental Research Funds for the Central Universities, Peking University (7100604896), and the China Manned Space Program (CMS-CSST-2025-A09). 
Luis C. Ho is supported by the National Science Foundation of China (12233001) and the China Manned Space Program (CMS-CSST-2025-A09). 
Yuan Bian and Min Du acknowledge the National Natural Science Foundation of China (No. 12573010), the Fundamental Research Funds for the Central Universities (No. 20720230015), and the Natural Science Foundation of Xiamen China (No. 3502Z202372006).
Zhao-Yu Li is supported by the National Natural Science Foundation of China under grant No. 12233001, by the National Key R\&D Program of China under grant No. 2024YFA1611602, and by a Shanghai Natural Science Research Grant (24ZR1491200).
Jing Wang thanks support of research grants from  Ministry of Science and Technology of the People's Republic of China (NO. 2022YFA1602902), National Science Foundation of China (NO. 12233001, 8200906879), and the China Manned Space Project.
\end{acknowledgments}

\begin{contribution}

Ming-Yang Zhuang discovered the system, processed the imaging data, conducted imaging decomposition, spectral modeling, and TNG50 analysis, developed the scientific interpretation, and wrote the initial draft. Jinyi Shangguan led the SED modeling. Yuan Bian gathered data from TNG50. Yue Shen and Luis C. Ho revised the manuscript. All authors contributed to the scientific interpretation and the final presentation of the manuscript.


\end{contribution}

%
\facilities{HST (ACS), JWST (NIRCam), SDSS}

\software{\texttt{astropy} \citep{2013A&A...558A..33A, 2018AJ....156..123A, 2022ApJ...935..167A},
\texttt{drizzlepac} \citep{2012drzp.book.....G}
\texttt{GALFITM} \citep{Haussler+2013MNRAS}, 
\texttt{Matplotlib} \citep{Hunter2007}, 
\texttt{Numpy} \citep{Harris2020}, 
\texttt{photutils} \citep{photutils}, 
\texttt{pPXF} \citep{pPXF},
\texttt{PSFEx} \citep{Bertin2011ASPC},
\texttt{pypher} \citep{pypher}
}



\bibliography{sample701}{}
\bibliographystyle{aasjournalv7}



\end{document}